\documentclass[12pt]{iopart}
\usepackage{hhline}
\usepackage{graphicx}
\usepackage{enumerate}
\usepackage[english]{babel}

\newcommand{\evec}{\textbf{e}}
\newcommand{\Fvec}{\textbf{F}}

\newcommand{\vvec}{\textbf{v}}
\newcommand{\rvec}{\textbf{r}}

\newcommand{\Mm}{\textbf{M}}

\renewcommand{\P}{\mathcal{P}}
\newcommand{\R}{\mathcal{R}}
\newcommand{\F}{\mathcal{F}}
\newcommand{\La}{\mathcal{L}}
\newcommand{\M}{\mathcal{M}}
\newcommand{\I}{\mathcal{I}}
\newcommand{\Q}{\mathcal{Q}}
\renewcommand{\S}{\widetilde{S}}

\newcommand{\al}{\alpha}
\newcommand{\en}{\epsilon}
\newcommand{\ta}{\theta}
\newcommand{\ti}{\theta_i}

\newcommand{\D}{\Delta}

\newcommand{\tM}{\tilde{M}}

\begin{document}

\title[Trebuchet with friction]
{The swinging counterweight trebuchet\\
On internal friction}

\author{E Horsdal}
\address{Department of Physics and Astronomy, Aarhus University,
DK-8000 Aarhus C, Denmark}
\ead{horsdal@phys.au.dk}
\begin{abstract}
Mechanical energy is lost to friction during a shot with a trebuchet.
The losses are mainly due to sliding friction at the bearings for the 
throwing arm and at the hinge for the swinging counterweight, 
but the aerodynamic force on the sling also contributes. 
Generalized forces for these sliding and aerodynamic frictions are derived and included
in the equations for the internal movement of the engine.
The equations are solved by the use of perturbation theory and calculated losses are compared 
with results from an experimental engine of small dimensions.
Scaling to full-size trebuchets is discussed.
\end{abstract}
%
%
\section{Introduction}
A swinging counterweight trebuchet is powered solely by gravity, 
but internal forces to constrain the internal movement are also present.
Gravity is a conservative force and the constraining forces do no work, 
so total mechanical energy is conserved.
However, mechanical energy is lost in practice due to frictional forces on 
rotating shafts and air drag on the rapidly moving sling carrying the projectile.
Although experiments have shown small losses by friction~\cite{ref:EPJ}, 
these cannot be ignored in a detailed analysis.

The generalized form of each frictional force is derived by the use of dissipation 
functions and added to the equations for the internal movement.
The forces from sliding friction prevents the equations from being 
solved directly by standard numerical methods,
but for small losses, such as those found in well-designed trebuchets, this problem 
can be overcome by perturbation theory and a few iterative solutions
to ensure self consistency.

The experimental results~\cite{ref:EPJ} were obtained by the use of a smaller engine 
equipped with motion sensors that make it possible to determine mechanical energies and losses.
Calculated losses underestimate the experimental results~\cite{ref:EPJ}, and degrees of 
freedom not included in the analysis are believed to be the explanation for this.
The results are extrapolated by scaling to large trebuchets inspired by historical renderings.
The experimental throwing arm, however, has an unusual mass distributions due to the 
sensors and their mountings so it was redesigned, but very capable engines of large
dimensions followed after this amendment.
\section{The trebuchet and friction}
\begin{figure}[htb]
	\centering	
	\includegraphics[width=1\textwidth]{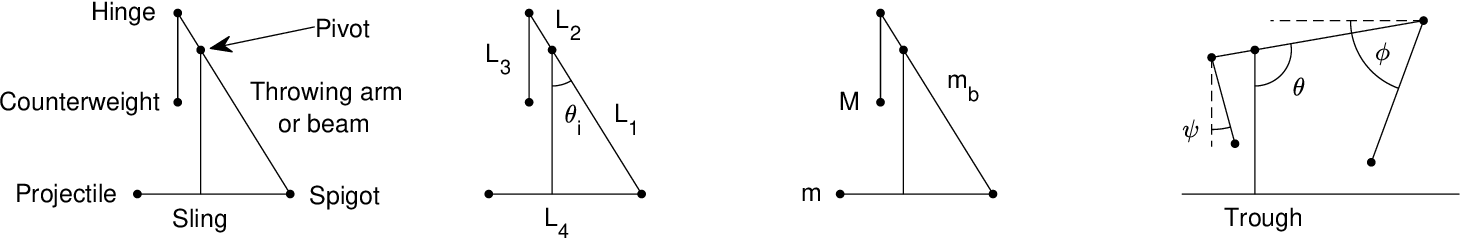}
		\caption{Trebuchet. Height of pivot is~$H=L_1\cos\ti$.}
		\label{fig:Treb}
\end{figure}

A detailed description of a trebuchet with swinging counterweight
can be found in~\cite{ref:EH}.
Here, we give only a short summery for convenience based on~\fref{fig:Treb},
which shows four schematic diagrams of the same trebuchet.
The various moving parts of the engine are identified in the diagram to the left. 
The next shows the long~$L_1$ and short~$L_2$ segments of the throwing arm, 
the arm for the counterweight~$L_3$, and the length of the sling~$L_4$. 
Terms used for the masses of projectile, counterweight and beam are given in the third, 
and generalized angular coordinates are seen at last for a particular instant of time 
after the projectile has been lifted from the trough.
These coordinates describe the movement of throwing arm~$\theta$, counterweight~$\psi$,
and sling~$\phi$.
The initial values are~$(\theta,\psi,\phi)_i=(\ti,0,0)$. 

A shot runs through three phases.
In phase~I, the projectile slides in a trough at the base of the engine,
but is eventually lifted and this marks the start of phase~II. 
This lasts until the projectile is released from the sling and flies towards the target,
which is located to the left.
The engine comes to rest during phase~III.

A pouch with two ropes forms a sling that carries the projectile, and a release mechanism 
comprising a ring and a spigot, often in the form of a hook, controls the opening of the pouch. 
One rope is tied to the ring, which is hung on the spigot from which it can slide off. 
The other is permanently tied to the throwing arm next to the spigot.
                                            
The position of the projectile in phase~II is
\begin{eqnarray}\nonumber
	\rvec=H\evec_y+
						L_1(\sin\ta\evec_x-\cos\ta\evec_y)-
						L_4(\cos\phi\evec_x+\sin\phi\evec_y)
\end{eqnarray}
where~$\evec_x$ and~$\evec_y$ are unit vectors in horizontal and vertical 
directions, respectively.
Similar expressions exist for the centers of mass of counterweight and pivoting beam,
and for phase~I.
This allows all potential energies to be written as functions of the generalized
coordinates~$(\ta,\psi,\phi)$, and the sum is~$U$.
Velocities follow from positions by differentiation with respect to time,
and this allows all kinetic energies to be written as functions 
of~$(\ta,\psi,\phi)$ and the derivatives~$(\dot\ta,\dot\psi,\dot\phi)$.
The sum is~$T$.
The three equations for the internal movement are derived from the generating
Euler-Lagrange equations
\begin{eqnarray}\label{eq:lagrange_equation}
	\frac{d}{dt}\frac{\partial\La}{\partial\dot q}-
		\frac{\partial\La}{\partial q}=0,
\end{eqnarray}
where~$q$ represents~$\ta$,~$\psi$ or~$\phi$ and~$\La=T-U$ is the Lagrange function.

Mechanical energy is lost to friction during the internal movement
and this is not taken into account in~\eref{eq:lagrange_equation}.
Non-conservative friction forces can be included in the analysis when they are 
expressed as generalized forces~$\R$ that depend on the generalized coordinates 
and their derivatives.
These forces are added on the right hand side of~\eref{eq:lagrange_equation}.
The calculation of the generalized forces is facilitated by dissipation 
functions~$\F$ introduced in~\sref{sec:Fric}, and in terms of these they are
\begin{eqnarray}\label{eq:GF}
	\R_q=-\frac{\partial\F}{\partial\dot q}~.
\end{eqnarray}
\section{Dissipation functions~$\F$ and generalized forces $\R$}\label{sec:Fric}
The sliding friction losses at the bearings are most often significantly larger 
than the aerodynamic loss relating to the motion of the sling.
The dominant sliding friction is discussed first and then air drag.
\subsection{Sliding friction}
The magnitudes of the time-dependent reaction forces at the bearings for the
beam and counterweight shafts are~$F_R=|\Fvec_R|$ and~$F_H=|\Fvec_H|$, respectively, 
and the sliding speeds are~$R_R\dot\ta$ and~$R_H(\dot\ta-\dot\psi)$, 
where~$R_R$ and~$R_H$ are radii of the shafts.
Two bearings, each carrying half weight, and the standard model for friction 
give the rate of heat generation
\begin{eqnarray}\label{eq:Power}
	P_f=\mu_RF_RR_R|\dot\ta|+\mu_HF_HR_H|\dot\ta-\dot\psi|,
\end{eqnarray}
where~$\mu_R$ and~$\mu_H$ are empirical friction coefficients.
The internal forces in~\eref{eq:Power} were discussed in~\cite{ref:EH_Forces}.
The model assumes two flat surfaces that slide against each other with
a certain area of contact, and this is problematic for two cylinders of 
not exactly the same radii.
However, wood is elastic so a finite area of contact forms, 
and what is more important,~\Eref{eq:Power} is independent of this area.
The model is therefore assumed to be applicable.
Each term in~\eref{eq:Power} is proportional to the appropriate radius, 
so this should be as small as possible to limit losses, but large enough for 
the shaft to carry the dynamic weight of the moving counterpoise or throwing arm. 
\begin{enumerate}
\item Generalized forces at bearings for beam shaft.\\
	The force~$\Fvec_R$ has the form~$-h\vvec/v$, where~$v=|\vvec|$ is the speed
	at contact and~$h=\mu_RF_R$ is positive and constant. 
	The dissipation function for a friction force of this form is given by
	\begin{eqnarray}\nonumber
		\F=\int_0^vh(v')dv'
	\end{eqnarray}
	so
	\begin{eqnarray}\nonumber
		\F=\mu_RF_Rv=\mu_RF_RR_R|\dot\ta|.
	\end{eqnarray}
	The generalized force follows from~\eref{eq:GF}, and with the sign function sng it reads
	\begin{eqnarray}\nonumber
		\R_{\theta1}	= -\mu_RF_RR_R\cdot
			\left\{ 
				\begin{array}{rcl}
					-1 & \mathrm{for} &\dot\ta<0 \\
					0  & \mathrm{for}	&\dot\ta=0 \\
					1  & \mathrm{for}	&\dot\ta>0
				\end{array}
			\right.
			\quad= -\mu_RF_RR_R\cdot\mathrm{sgn}(\dot\ta).
	\end{eqnarray}
\item Generalized forces at hinge.\\
	The dissipation function is in this case
	\begin{eqnarray}\nonumber
		\F=\mu_HF_Hv=\mu_HF_HR_H|\dot\ta-\dot\psi|,
	\end{eqnarray}
	so one finds two generalized forces
	\begin{eqnarray}\nonumber
		\R_{\theta2}	= -\mu_HF_HR_H\cdot\mathrm{sgn}(\dot\ta-\dot\psi)
	\end{eqnarray}
	and
	\begin{eqnarray}\nonumber
		\R_{\psi}	= -\mu_HF_HR_H\cdot\mathrm{sgn}(\dot\psi-\dot\ta).
	\end{eqnarray}
\end{enumerate}
\subsection{Aerodynamic friction}
The aerodynamic force on the sling is modeled by the standard form
\begin{eqnarray}\label{eq:F_air}
	\Fvec = -\frac{1}{2}\rho_a CA v^2\frac{\vvec}{v},
\end{eqnarray}
where~$\rho_a$ is air density,~$C$ an aerodynamic constant,~$A$ an appropriate 
area representing the sling and~$\vvec$ the projectile velocity.
The value of~$C$ is close to~1/4 for objects and speeds like the present, 
and for the aerodynamic cross section we use 
\begin{eqnarray}\nonumber
	A=\al\cdot\pi\left(\frac{3m}{4\pi\rho_s}\right)^{2/3},
\end{eqnarray}
where~$\rho_s\simeq2700$kg/m$^3$ is the assumed density of stone.
The factor~$\al=2$ is included such that~$A$ is twice the cross section of the 
projectile to simulate the aerodynamic cross section of the pouch.

The dissipation function for the force in~\eref{eq:F_air} is
\begin{eqnarray}\nonumber
	\F=\frac{1}{2}\rho_a CA\int_0^vv'^2dv'=\frac{1}{6}\rho_a CAv^3,
\end{eqnarray}
with~$v^3$ given by
\begin{eqnarray}\nonumber
	v^3	=	\left(
						L_1^2\dot\ta^2+L_4^2\dot\phi^2-2L_1L_4\sin(\theta-\phi)\dot\ta\dot\phi
				\right)^{3/2}.
\end{eqnarray}
Two generalized forces now follow. The first is
\begin{eqnarray}\nonumber
	\R_{\theta3}	= -\frac{\partial\F}{\partial\dot\ta} 
						= -\frac{1}{6}\rho_a CA\frac{\partial v^3}{\partial\dot\ta} 
						= -\frac{1}{2}\rho_a CAL_1vv_\ta,
\end{eqnarray}
where
\begin{eqnarray}\nonumber
	v_\ta	= L_1\dot\ta-L_4\sin(\theta-\phi)\dot\phi,
\end{eqnarray}
and the second
\begin{eqnarray}\nonumber
	\R_{\phi}	= -\frac{\partial\F}{\partial\dot\phi} 
						= -\frac{1}{6}\rho_a CA\frac{\partial v^3}{\partial\dot\phi} 
						= -\frac{1}{2}\rho_a CAL_4vv_\phi,
\end{eqnarray}
where
\begin{eqnarray}\nonumber
	v_\phi	= L_4\dot\phi-L_1\sin(\theta-\phi)\dot\ta.
\end{eqnarray}
\section{Equations of internal movement with friction}\label{sec:EOM_FRIC}
The equations of motion including sliding and aerodynamic friction read
\begin{eqnarray}\label{eq:EOM_fric}
	\frac{d}{dt}\frac{\partial\La}{\partial\dot q}-
		\frac{\partial\La}{\partial q}=\R_q,
\end{eqnarray}
where~$q$ represents~$\ta$,~$\psi$ or~$\phi$, 
and~$\R_\ta=\R_{\theta1}+\R_{\theta2}+\R_{\theta3}$.

The speed of pouch and projectile is small in phase~I, so aerodynamic friction 
can be neglected here and treated only in phase~II, but sliding friction is included 
in all three phases of a shot.
Phase~III is considered, because the kinetic energy remaining in the engine after release 
must be dissipated before a new shot can be prepared.
The lowest possible friction is therefore preferable for a single shot, but may be
a limiting factor for the frequency of shots.

The two generalized aerodynamic force terms~$\R_{\theta3}$ and~$\R_{\phi}$ 
do not change the character of the equations of motion, which can still be cast 
into the form of six coupled, first-order differential equations.
In matrix notation 
\begin{eqnarray}\label{eq:EOM}
	\frac{d}{dt}\{\theta,\psi,\phi,\dot\ta,\dot\psi,\dot\phi\}^T=
	\Mm\cdot\{\dot\ta,\dot\psi,\dot\phi\,f_1,f_2,f_3\}^T,
\end{eqnarray}
where~$\{\dots\}^T$ is the transpose of~$\{\dots\}$,~$\Mm=\Mm(\theta,\psi,\phi)$ 
is a~$6\times6$ matrix that depends on the angular coordinates, but not their derivatives, 
and~$f_i=f_i(\theta,\psi,\phi,\dot\ta,\dot\psi,\dot\phi)$. 
The six equations can be solved by standard numerical techniques, and this 
determines the total mechanical energy~$E_{tot}$ as a function of time.
The loss of mechanical energy~$Q(t)=E_{tot}(0)-E_{tot}(t)$ must equal the 
work done by the aerodynamic force~$\Fvec$ in~\eref{eq:F_air}
\begin{eqnarray}\nonumber
 	Q(t)=-\int_0^t\Fvec\cdot\vvec dt=
	\frac{1}{2}\rho_a CA\int_0^t v^3dt,
\end{eqnarray}
and this consistency test was performed affirmatively within numerical uncertainty.

The equations of motion are changed more fundamentally by the inclusion of 
the sliding friction forces~$\R_{\theta1}$,~$\R_{\theta2}$ and~$\R_{\psi}$,
because they depend on the reaction forces~$F_R$ and~$F_H$, which in turn 
depend on the angular accelerations~$(\ddot\ta,\ddot\psi,\ddot\phi)$ in a way that
prevents the equations from being cast into the form~\eref{eq:EOM} and solved immediately.
However, if the losses are small, the equations can be solved iteratively as they stand:
The motion is first solved with~$F_R=F_H=0$.
This allows~$F_R(t)$ and~$F_H(t)$ to be calculated to lowest order from the 
unperturbed motion.
The generalized forces~$\R_{\theta1}$,~$\R_{\theta2}$ and~$\R_{\psi}$ are then 
calculated and treated as explicitly known source terms,
which become part of the functions~$f_i$ in~\eref{eq:EOM}.
The motion is thereafter solved with this first estimate of the friction 
and this allows an improved estimate to be derived.
The procedure converges in a few steps for small losses, and can be checked for 
conservation of total energy by comparing the loss of mechanical energy 
calculated from the motion with the work done by the sliding friction forces
\begin{eqnarray}\label{eq:Q_t}
	Q(t)=
	\mu_RR_R\int_0^tF_R(t)|\dot\ta|dt+\mu_HR_H\int_0^tF_H(t)|\dot\ta-\dot\psi|dt
\end{eqnarray}
as was pointed out already for the aerodynamic losses.
\section{Results and comparison with experiment}
The internal movement of an experimental trebuchet was recently determined by the use of
rotation sensors to measure the three angles~$\theta$,~$\psi$ and~$\phi$~\cite{ref:EPJ}.
Positions and velocities of the projectile and all parts of the engine could be determined 
as functions of time during a shot from these measurements. 
All potential and kinetic energies and all forces could be determined as well.
The experiments definitely show loss of mechanical energy during phases~I and~II of a shot, 
and about one half can be attributed to friction.
It was also speculated~\cite{ref:EPJ} that the other half could be found in degrees of 
freedom that are not included in the analysis.
The mechanical energy remaining in the engine after release is 
lost over several oscillation periods of the throwing arm and counterweight.

Numerical values of the design parameters for the experimental trebuchet, 
which includes lengths and masses, are given in the first columns of~\Tref{tab:Param}.
The respective radii~$R_R$ and~$R_H$ of the beam and hinge shafts are also included, and
the length~$L_R$ of the beam shaft from bearing to bearing is important too and included 
because it is a determining factor for the strength of the shaft.
\begin{table}[tbh]\footnotesize   
	\centering
		\begin{tabular}{cccc|ccc||ccc|cc|cc}
			\multicolumn{4}{c|}{Lengths} 	& \multicolumn{3}{c||}{Masses} & \multicolumn{3}{c|}{Beam} & 
			\multicolumn{2}{c|}{Hinge} & \multicolumn{2}{c}{Sling} 																										\\\hline
			$L_1$ & $L_2$ &	$L_3$ & $L_4$ & $M$		& $m$ 	& $m_b$ & $R_R$	& $\mu_R$	& $L_R$	& $R_H$	& $\mu_H$	& $A$			& $C$	\\
			cm    & cm    & cm    & cm 		& kg 		& kg 		& kg 		& cm		&  				& cm		& cm		& 				& cm$^2$	& 		\\\hline
			97.5	& 25.0	& 51.5	& 87.0	& 53.9 	& 0.717 & 4.86  & 1.7		& 0.35 		& 34		& 0.25	&	0.35		& 100			& 0.25	
		\end{tabular}
	\caption{Parameters of experimental trebuchet with $\ti=31.5^\circ$ and $\D U$=204J.
	$I_B$=1.85kgm$^2$ and~$L_{CM}$=46.5cm.}
	\label{tab:Param}
\end{table}
Assumed values of friction coefficients~$\mu_R$ and~$\mu_H$, the aerodynamic cross section~$A$
of the sling, and the drag coefficient~$C$ are also given.
The moment of inertia of the throwing arm~$I_B$ and the position of its center of mass~$L_{CM}$ 
relative to the pivot were determined as discussed in~\cite{ref:EPJ}. 
These two beam parameters include contributions from two relatively heavy metallic fittings 
at the long and short ends of the throwing arm. 
One holds and supports the spigot and the other is a hinge bracket for the counterweight. 
Fittings for the rotation sensors and reinforcements near the pivoting shaft also influence the two 
parameters.
\subsection{Convergence of iterative procedure}
\begin{figure}[htb]
	\centering	
	\includegraphics[width=0.85\textwidth]{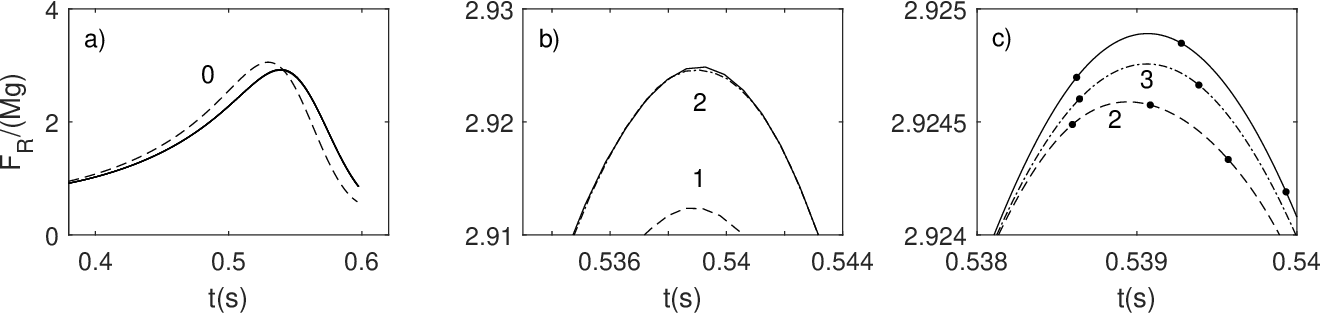}
		\caption{Reaction force~$F_R$ at fulcrum in units of~$Mg$.}
		\label{fig:Forces_Conv}
\end{figure}

\Fref{fig:Forces_Conv} 
shows normal reaction forces at the bearings for the throwing arm that illustrate the convergence 
of the repeated solutions discussed in~\sref{sec:EOM_FRIC}.
The calculations were done with the parameters of the experimental trebuchet in~\tref{tab:Param}.
The first approximation is derived from the internal movement without friction and is iteration~0.
It is shown in~\fref{fig:Forces_Conv}a
by the broken curve, and the final result after four iterations is the full curve.
Iterations~1,~2 and the final are shown in~\fref{fig:Forces_Conv}b near the maximum, but 
iteration~2 is difficult to distinguish from the final result. 
\Fref{fig:Forces_Conv}c is zoomed in even stronger on the maximum, and here iterations~2 and~3 
are both seen.
Even though convergence has been achieved already at the second iteration in the given example, 
iteration~4 is in general taken to be the final converged result.

\begin{figure}[htb]
	\centering	
	\includegraphics[width=0.70\textwidth]{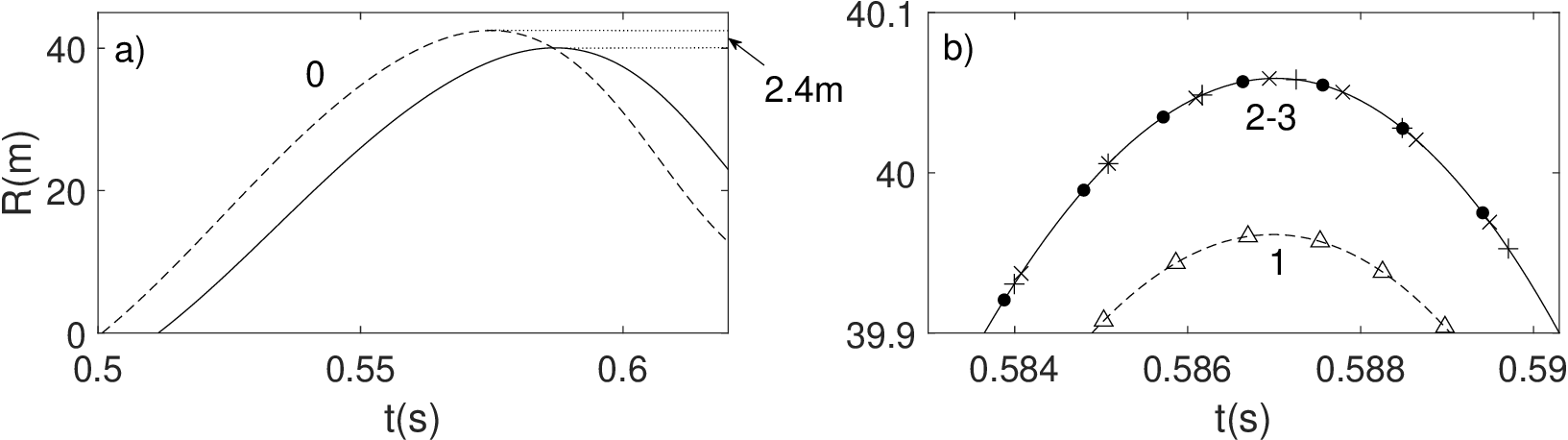}
		\caption{Range~$R$ in vacuum vs release time.}
		\label{fig:Range_Conv}
\end{figure}

The internal movement becomes a little slower and less violent when friction is present.
This was seen already in~\fref{fig:Forces_Conv}a and illustrated also in~\fref{fig:Range_Conv} 
by calculated projectile ranges in vacuum as functions of the time of release from the sling. 
The calculations were again done with the parameters in~\tref{tab:Param}.
Iteration~0 in which friction is absent and the final iteration are shown 
in~\fref{fig:Range_Conv}a as broken and full curves, respectively.
Friction delays the release for longest range by~$\simeq10$ms and reduces the range 
by~2.4m or~$\simeq5.6\%$.
The mechanical energy~$E_m$ gained by the projectile is likewise reduced 
by~$\simeq5.5\%$ to~154J from~163J.
Approximately equal relative reductions are expected because release energy is approximately
proportional to vacuum range for good shots.
Intermediate iterations are shown in~\fref{fig:Range_Conv}b.
Iteration~1, shown by a broken curve and triangles, suppresses range too much, 
but the next iterations including iteration~4 are all converged.
\subsection{Rates of mechanical energy loss}
\begin{figure}[htb]
	\centering	
	\includegraphics[width=0.75\textwidth]{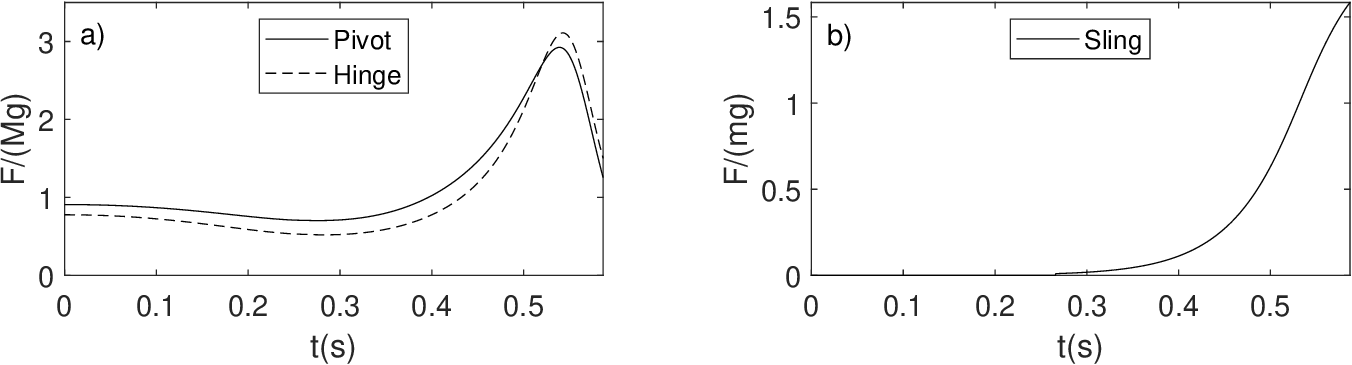}
		\caption{
		a) Reaction forces at pivot and hinge.
		b) Aerodynamic force on sling.}
		\label{fig:Forces}
\end{figure}

Normal reaction forces at pivot and hinge are shown in~\fref{fig:Forces}a
as functions of time from start at~$t=0$ and until release for maximum vacuum range at~$t=0.587$s, 
once more with the parameters of the experimental trebuchet in~\tref{tab:Param}.
The forces are smaller than~$Mg$ during most of the shot while the counterweight is falling almost
straight down, but they both rise sharply and go through maxima near~$3Mg$ 
shortly before release when this fall is interrupted and 
the counterweight goes into its final oscillatory motion.
The force at the hinge is the smallest of the two at start, but it ends a little higher.
The aerodynamic force on the sling shown in~\fref{fig:Forces}b is very small during most
of the shot when the projectile speed is still small, but rises sharply towards release 
where it becomes somewhat larger than~$mg$.

\begin{figure}[htb]
	\centering	
	\includegraphics[width=0.50\textwidth]{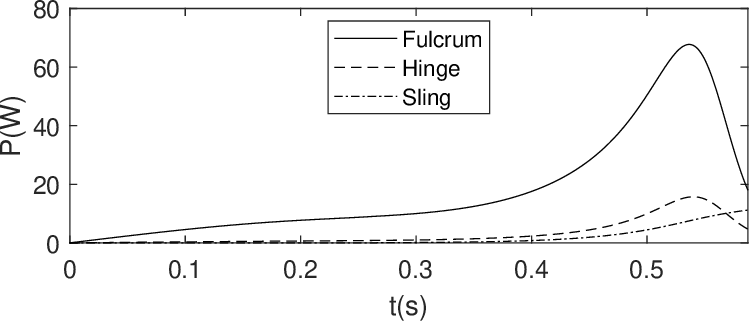}
		\caption{
		Powers at fulcrum, hinge and sling until release for maximum range.}
		\label{fig:Powers}
\end{figure}

The forces in~\fref{fig:Forces} and friction coefficients in~\tref{tab:Param} lead
to the rates of losses shown in~\fref{fig:Powers}.
The losses at the pivot are clearly much larger than at the hinge even though the friction forces 
are almost equal and the angular rotation speeds differ only little.
The explanation is found in the different radii of the pivoting shafts. 
Aerodynamic drag on the sling contributes only little to the accumulated loss, but
becomes significant shortly before release.
The sum of powers reaches a maximum of~91.5W at~0.537s and here drag contributes 
by~9\%.
At release, however, the power is~33.9W and drag now contributes by~33\%. 
\subsection{Loss of mechanical energy and efficiency}\label{sec:LME}
Calculated friction losses are obtained by integrating the rates in~\fref{fig:Powers} over time
and results are listed in~\tref{tab:Losses}.
\begin{table}[htb]\footnotesize
	\centering
		\begin{tabular}{cc|cc|cc||cc|cc||ccc}
			\multicolumn{2}{c|}{Fulcrum} 		& \multicolumn{2}{c|}{Hinge} 				& \multicolumn{2}{c||}{Sling} 		&
			\multicolumn{2}{c|}{Total}			& \multicolumn{2}{c||}{Total}				& \multicolumn{3}{c}{Efficiency}	\\
			\multicolumn{2}{c|}{Wood-wood}	& \multicolumn{2}{c|}{Steel-steel}	& \multicolumn{2}{c||}{Air} 			&
			\multicolumn{2}{c|}{Absolute}		& \multicolumn{2}{c||}{Relative}		&	\multicolumn{3}{c}{$\en$}				\\
			\multicolumn{2}{c|}{$Q_P$(J)} 	& \multicolumn{2}{c|}{$Q_H$(J)} 		& \multicolumn{2}{c||}{$Q_A$(J)}	&
			\multicolumn{2}{c|}{$Q$(J)}			&	\multicolumn{2}{c||}{$Q/\D U$}																			\\
			Cal		& Exp 	& Cal	& Exp	& Cal	& Exp	& Cal 	& Exp	& Cal 	& Exp	& Cal, ideal	& Cal 		& Exp  			\\\hline
			10.7	& 11.8	& 1.8	& 2.0	& 1.0	& 0.9	& 13.5	& 33	& 6.6\%	& 16\%	& 80.4\%		& 75.5\%	& 68.8\%
		\end{tabular}
	\caption{Losses at pivot, hinge and sling. 
	Total losses.
	Efficiencies, ideal and with losses.	$\D U=204J$.
	Cal: Present calculations. Exp: Experimental results from~\cite{ref:EPJ}.}
	\label{tab:Losses}
\end{table}
The calculated loss at the fulcrum amounts to~10.7J and this should be compared with the~11.8J 
found experimentally~\cite{ref:EPJ}.
The two values are found by using the same model for friction, so the insignificant 
difference between them reflects agreement between calculated and experimental internal movements.
Calculated and experimental losses at the hinge also agree, but are much smaller than at the pivot.
Wood is used for the pivoting shaft of the beam and its bearings, so wood slides against wood.
The materials used for the counterweight arm and its hinge to the beam are stainless steel.
The friction coefficients for wood and steel are assumed to be equal, but steel is much stronger
than wood, so the hinge was designed with a much smaller radius than that of the wooden shaft, 
and this explains the reduced losses.
Calculated and experimental aerodynamic losses also agree and they contribute the least.

The total calculated loss of~13.5J is the sum of the friction losses, 
but the total experimental loss of~33J is more than twice the sum of the experimental frictions,
which is~14.7J.
An increase of the empirical friction coefficient~$\mu_R$ would narrow the gab, but an elimination
requires~$\mu_R\simeq0.7$, which is considered unrealistic and also leads to 
an impossible long-term increase of apparent mechanical energy as stressed in~\cite{ref:EPJ}.
The energy that has not been accounted for near or shortly after the release has most likely migrated 
into one or more ignored degrees of freedom, of which there may be many:
Deformation and movement of the trestle, stretching and bending of the throwing arm, 
expansion of the sling, etc.

The difference of potential energies of the engine in initial and final positions is
\begin{eqnarray}\label{eq:DU}
	\D U=(ML_2-m_bL_{CM})g(1+\cos\ti).
\end{eqnarray}
This is the largest amount of mechanical energy that can be transferred to the projectile.
The efficiency~$\en$ of the engine is the fraction of~$\D U$ that is actually transferred, so
\begin{eqnarray}\nonumber
	\en=\frac{E_m}{\D U},
\end{eqnarray}
where~$E_m$ is the mechanical energy gained by the projectile at release.
The calculated ideal efficiency in the absence of losses is~80.4\%, see~\tref{tab:Losses}.
The total loss is distributed on the engine and the projectile, 
and the fraction carried by the projectile reduces the calculated efficiency of the engine to~75.5\%.
The somewhat lower experimental efficiency of~68.8\% includes the losses that are not accounted for
with certainty.
\subsection{Loss of range}
The experiments~\cite{ref:EPJ} include field measurements of longest range~$R_f$.
This quantity was determined after many shots with varying spigot angle to find the best setting, 
and many shots under identical conditions as far as possible to obtain a standard deviation.
Numerical values are given in~\tref{tab:Range}.
\begin{table}[htb]\footnotesize
	\centering
		\begin{tabular}{c|cc|ccc}
			\multicolumn{1}{c|}{Field} 				& \multicolumn{5}{c}{Calculation} 		\\\cline{2-6}
			\multicolumn{1}{c|}{measurement} 	& \multicolumn{2}{c|}{no friction} & 
			\multicolumn{3}{c}{friction} 																						\\\hline
			$R_f$ 				& $R_v$ & $R_a$ & $R_v$ & $R_v^f$ & $R_a^f$ 							\\
			m 						& m 		& m 		& m 		& m 			& m 										\\\hline
			34.4$\pm$1.5	& 42.8	& 39.5	& 42.8	& 40.4 		& 37.3 		
		\end{tabular}
	\caption{Calculated and experimental ranges with standard deviations.}
	\label{tab:Range}
\end{table}
The experimental range was compared in~\cite{ref:EPJ} with a calculated range~$R_a$ derived 
from the vacuum ranges~$R_v$ corrected for aerodynamic drag along the ballistic trajectory 
to the target. 
These ranges are also included in~\tref{tab:Range}.
Here, comparison is made with the range~$R_a^f$, which is derived from the vacuum range~$R_v$ 
first corrected for friction~$R_v^f$ and then for aerodynamic drag.
The prediction of~$R_a$ without friction in column three lies~3.4 standard deviations over
the experimental field value~$R_f$, 
but the deviation is reduced by almost a factor of two to~1.9 standard deviations 
with the inclusion of friction as in~$R_a^f$ shown in the last column.
The remaining difference agrees with an amount of energy not accounted for as mentioned 
in~\sref{sec:LME}.
\section{Discussion and scaling}
The current experimental results for a small trebuchet are in this section scaled to the size 
of large historical engines, but this involves complications related to the specific 
experimental design, where:
\begin{enumerate}	
	\item~\label{page:L_3} 
		The counterweight is made of stainless steel, which allows its center of mass to fall
		allmost to the level of the trough due to the high density of steel.
		
		\vspace{10pt}
		Historical trebuchets with swinging counterweights are usually depicted with a wooden 
		box filled with materials such as stones and wet soil.
		Although these materials are heavy, they do not have the high density of steel so the 
		arm~$L_3$ for the counterweight becomes too long if it is scaled like other lengths.
	\item		 
		The counterweight is hinged to the beam by a metallic axle of small diameter. 
		
		\vspace{10pt}
		The counterweight box in historical engines is hinged to the throwing arm by a shaft 
		made of wood instead of steel, and most often, this shaft appears to have a diameter 
		similar to or slightly smaller than the shaft that carries the throwing arm, 
		see~\cite{ref:DB} and plates~26,~29 and~30 of~\cite{ref:VdH}.
		The sliding friction losses at the two shafts are then approximately equal.
	\item
		The two sides of the trestle that supports the throwing arm stand vertically
		with a constant distance between them, and this implies a long shaft for the beam.
		
		\vspace{10pt}
		The trestle must be sufficiently wide near the base to allow free 
		swing of the counterweight box, but the frame is often made narrower near the top,
		presumably to allow for a shorter shaft.
		An illustration is seen in figure~2 of~\cite{ref:DR}, which is an artistic 
		drawing of an engine intended to resemble the first ones used in England 
		early in the 13th century.		
	
	\item
		The throwing arm has an unusual mass distribution with its many fittings.	
		
		\vspace{10pt}
		The depicted throwing arms are tapered to varying degrees, but we assume a uniform 
		cylindrical arm  with a moment of inertia found by scaling the experimental value.
		This choice is made because the internal motion of the experimental and scaled engines 
		should be as comparable as possible, and among the parameters of the throwing arm, 
		the moment of inertia is the most important for the dynamics.
\end{enumerate}

With reference to point~(ii) and~(iii), the two shafts are assumed to be identical
and they must also be sufficiently strong to safely carry the dynamic weights during a shot 
while not allowing the sliding friction losses to become too great.
A small radius limits losses, but weakens strength, so the shafts must be short to compensate.
Not only the shafts, but also the throwing arm is exposed to stress.
The risk of breaking it must therefore be examined too.

We first discuss the scaling properties of the equations for the internal movement, 
then the strengths of the shafts and the throwing arm, and finally
determine the parameters of the throwing arm.
After this, scaling properties are used to extrapolate the small experimental design 
to full size engines.
\subsection{Scaling of equations of motion}\label{sec:Scaling_EOM}
The equations for the internal movement depend on the 
parameters identified in~\fref{fig:Treb}, but not only these.
The center of mass distance~$L_{CM}$ and the moment of inertia~$I_B$ both relate 
to the throwing arm, and are parameters of the equations, and when sliding friction 
is included the equations also depend on the radii~$R_R$ and~$R_H$ of the shafts.
The equations become dimensionless when divided term by term by~$MgL$, 
where~$L=L_1+L_2$ is the length of the throwing arm, and if all lengths in the equations 
are measured in units of~$L$, masses in units of~$M$, time in units of~$(L/g)^{1/2}$ 
and small aerodynamic terms 
are left out, the dimensionless equations are scale invariant.
\subsection{Strengths of shafts and throwing arm}\label{sec:Sst}
The stress~$\sigma$ of the shaft that carries the throwing arm is related to its 
strain~$S$ by
\begin{eqnarray}\nonumber
	\sigma=\M_eS,
\end{eqnarray}
where $\M_e$ is Young's modulus of elasticity for the type of wood being used.
The modulus of rupture is the upper limit for the stress~$\sigma$ beyond which 
the wood breaks.
This is often near~1\% of~$\M_e$, so~$S$ must be somewhat smaller than~1\%.
The strain~$S$ at a given position along the shaft equals its curvature at that
position times its radius. 
For simplicity and not to underestimate the strain, we assume that the full load 
is applied at the middle of the shaft, where the curvature is largest.
During loading in preparation for a shot, the strain of the shaft goes through a 
maximum when the throwing arm is quasi-static in horizontal position.
At this point, the moment of force applied to the arm with respect to the pivot equals zero 
so~$MgL_2=M_sgL_1+m_bgL_{CM}$, where~$M_sg$ is the perpendicular force applied at the spigot. 
The total force at the middle of the shaft is then~$(1+L_2/L_1)Mg$ when~$m_b\ll M$, and the 
strain is
\begin{eqnarray}\nonumber
	S_{\mathrm{load}}=R_R\frac{\tM gL_R}{4\M_e\I}
	\quad\mathrm{with}\quad 
	\tM=\left(1+\frac{L_2}{L_1}\right)M
	\quad\mathrm{and}\quad        
	\I=\frac{\pi}{4}R_R^4,
\end{eqnarray}
where~$\I$ is the second moment of area for the circular cross section of the shaft
and~$L_R$ its length from bearing to bearing.
The parameters in~\tref{tab:Param} and~$\M_e=12$GPa lead to~$S_{\mathrm{load}}=0.12\%$ which is quite
safe.
However, in~\fref{fig:Forces} we saw that the reaction forces at fulcrum and hinge rise
to~$\simeq3$Mg shortly before the projectile is released, and this leads to a common estimate
for the two identical shafts
\begin{eqnarray}\nonumber
	S_{\mathrm{shot}}\simeq R_R\frac{3M gL_R}{4\M_e\I}=R_H\frac{3M gL_H}{4\M_e\I}.
\end{eqnarray}
Again with the parameters in~\tref{tab:Param}, we find~$S_{\mathrm{shot}}\simeq0.29\%$, 
which is also safe, but somewhat conservative, because the strain is near maximum 
only for a short time.
The estimate for the strain of the shafts to be used from here on is the compromise
\begin{eqnarray}\label{eq:shaft}
	S_{\mathrm{shaft}}\simeq R_R\frac{M gL_R}{2\M_e\I},
\end{eqnarray}
which gives~$S_{\mathrm{shaft}}=0.20\%$ with the parameters in~\tref{tab:Param}. 

The throwing arm is also strained. 
During loading, the maximum is found at the pivot when the arm is horizontal.
The force at the pivot is then~$\tM g$ and the strain is
\begin{eqnarray}\label{eq:arm}
	S_{\mathrm{arm}}=R_a\frac{\tM g}{\M_e\I}\frac{L_1L_2}{L_1+L_2}=R_a\frac{MgL_2}{\M_e\I}
	\quad\mathrm{with}\quad      
	\I=\frac{\pi}{4}R_a^4,
\end{eqnarray}
where~$R_a$ is the radius of the arm.
The bending load during a shot is the component of the reaction force~$\Fvec_R$ 
perpendicular to the beam.
It was discussed in~\cite{ref:EH_Forces} and there it rose to a maximum of~$1.60Mg$ 
shortly before release.
With the current experimental trebuchet the maximum is~$1.80Mg$, which is~45\% larger 
than~$\tM=1.25$Mg used in~\eref{eq:arm}, but it is larger for only~$\simeq0.10$s,
so we settle with~\eref{eq:arm}.
\subsection{Scaling at constant~$S_{\mathrm{shaft}}$}\label{sec:Scaling_S_shaft}
To limit friction losses and at the same time the strain~$S_{\mathrm{shaft}}$ given in~\eref{eq:shaft},
we select a relatively small common radius~$R_R=R_H$ for the shafts and this is possible when 
they are not too long. 
The common length is set at twice the diameter~$D$ of the throwing arm,~i.e.~$L_R=L_H=2D$, 
and it is therefore proportional to~$(m_b/L)^{1/2}$.
This and~\eref{eq:shaft} then implies~$S_{\mathrm{shaft}}\propto M^{3/2}/L^{7/2}$, which is
constant under scaling when masses vary like~$k^{7/3}$, 
where~$k$ is the scaling factor for the lengths in the equations of motion.
The lengths $D$ and~$L_R=L_H$, that do not enter the equations, 
then vary like~$(k^{7/3-1})^{1/2}=k^{2/3}$. 
Likewise, the strain of the throwing arm~$S_{\mathrm{arm}}$ in~\eref{eq:arm} varies
like~$k^{2/3+7/3+1-4\cdot2/3}=k^{4/3}$, and for the unit of energy~$MgL$ the variation is
like~$k^{7/3+1}=k^{10/3}$.
\subsection{Parameters of the throwing arm}\label{sec:Arm}
The experimental throwing arm has an unusual mass distribution with its relatively heavy fittings
and is replaced by a uniform cylinder made of wood to relate to historical engines. 
The equations for the internal movement depend on the throwing arm through the 
parameters~$L_1$,~$L_2$~$m_b$,~$L_{CM}$ and~$I_B$, but these do not fully determine it, 
because different distributions of the mass~$m_b$ exist for unchanged values of the five parameters.
The sections $L_1$ and~$L_2$ are found by scaling to keep proportions constant.
Among the remaining parameters~$m_b$,~$L_{CM}$ and~$I_B$, the inertia~$I_B$ is considered the most 
important for the dynamics, so this is also found by scaling,
but this choise implies modification of~$m_b$ and~$L_{CM}$.
The physical dimensions of the inertia~$I_B$ are mass multiplied by length squared, 
so with~\sref{sec:Scaling_EOM} and~\ref{sec:Scaling_S_shaft}, 
it scales like~$k^{7/3+2}=k^{13/3}$, while~$L_1$ and~$L_2$ scale like~$k$.

The sections~$L_1$ and~$L_2$ are now given by
\begin{eqnarray}\nonumber
	L_1=kL_1^e
	\qquad\mathrm{and}\qquad
	L_2=kL_2^e,
\end{eqnarray}
where~$L_1^e$ and~$L_2^e$ are the experimental values in~\tref{tab:Param}.
The modified distance from pivot to center of mass is then
\begin{eqnarray}\label{eq:LCMa}
	L_{CM}^m=\frac{1}{2}\left(L_1-L_2\right).
\end{eqnarray}
The modified mass~$m_b^m$ satisfies the equation
\begin{eqnarray}\label{eq:mba}
	\frac{1}{3}m_b^m\left(L_1^2-L_1L_2+L_2^2\right)=I_Bk^{13/3},
\end{eqnarray}
where~$I_B=1.85$kgm$^2$ is the experimental moment of inertia determined 
in~\cite{ref:EPJ}, and the diameter~$D$ of the cylinder follows from
\begin{eqnarray}\label{eq:Da}
	\rho\frac{\pi}{4}D^2(L_1+L_2)=m_b^m,
\end{eqnarray}
where~$\rho$ is the appropriate density of wood.
\subsection{Modified throwing arm and shortened arm for counterweight}
The design of the experimental trebuchet was given in~\tref{tab:Param}. 
Further details of the experimental throwing arm are listed in the 
first row of~\tref{tab:Param_Alt1}, and a few other parameters are repeated for convenience.
\begin{table}[htb]\footnotesize
	\center
		\begin{tabular}{c|cccc|cc|cc|cc}
			&\multicolumn{4}{c}{Throwing arm} & \multicolumn{2}{c|}{Shaft at fulcrum} 	
			& & &\multicolumn{2}{|c}{Losses at fulcrum}																											\\
			&$L_{CM}$		& $m_b$ & $I_B$ 	& $D$	& $R_R$ & $L_R$ & $L_3$ & $\D U$	& $Q$		& $Q/\D U$				\\
			&cm 				& kg		& kgm$^2$	& cm 	& cm 		& cm 		& cm 		& J				& J 		& \%							\\\hline
			Experimental	&46.5   		&	4.86	& 1.85  	&			&	1.7   & 34.0	& 51.5	& 204	& 10.7	& 5.25	\\	
			Modified 			&36.0   		&	7.21	& 1.85  	&	9.7	&	1.4   & 19.4	& 30.4 	&	198	&	8.82	& 4.45
		\end{tabular}
	\caption{Experimental and modified throwing arms and their shafts. 
	Arms for counterweight~$L_3$ and potential energies~$\D U$. 
	Absolute and relative sliding friction losses. 
	$L=L_1+L_2=1.225$m. $L_1/L_2=3.9$. 
	$S_{\mathrm{shaft}}=0.19\%$.}
	\label{tab:Param_Alt1}
\end{table}
The second row of the table shows the modified throwing arm with amended values for the 
center of mass distance~$L_{CM}$ and the weight~$m_b$.
These parameters are found by the use of~\eref{eq:LCMa} and~\eref{eq:mba} with~$k=1$, 
but other parameters of the arm are unchanged including the moment of inertia~$I_B$.
The diameter~$D$ of the throwing arm is given by~\eref{eq:Da}, and
the shaft that carries the arm is made shorter~$L_R=2D$ and slimmer such 
that~$S_{\mathrm{shaft}}$ equals~0.19\%.
The largest strain during loading of the modified throwing arm~$S_{\mathrm{arm}}$, which is given in~\eref{eq:arm}, 
turns out to be very small,~$S_{\mathrm{arm}}=0.006\%$.
During a shot it can be~45\% larger, see~\sref{sec:Scaling_S_shaft}, but is still small.

\Tref{tab:Param_Alt1} also lists a shorter modified length~$L_3$ of the arm for the 
counterweight, see comment (i) on page~\pageref{page:L_3}. 
This is given by~$L_2+L_3=\frac{2}{3}H$, where~$H=L_1\cos\ti$ is the height of the bearings.
A reduced difference of potential energies~$\D U$ due to the~15\% increase 
of the term~$m_bgL_{CM}$ in~\eref{eq:DU} is also given.
The calculated loss of mechanical energy at the fulcrum of the experimental engine 
is~10.7J which amounts to~5.25\% of~$\D U$.
The values with the modified throwing arm are smaller because of the smaller~$R_R$.

\begin{figure}[htb]
	\centering	
	\includegraphics[width=0.95\textwidth]{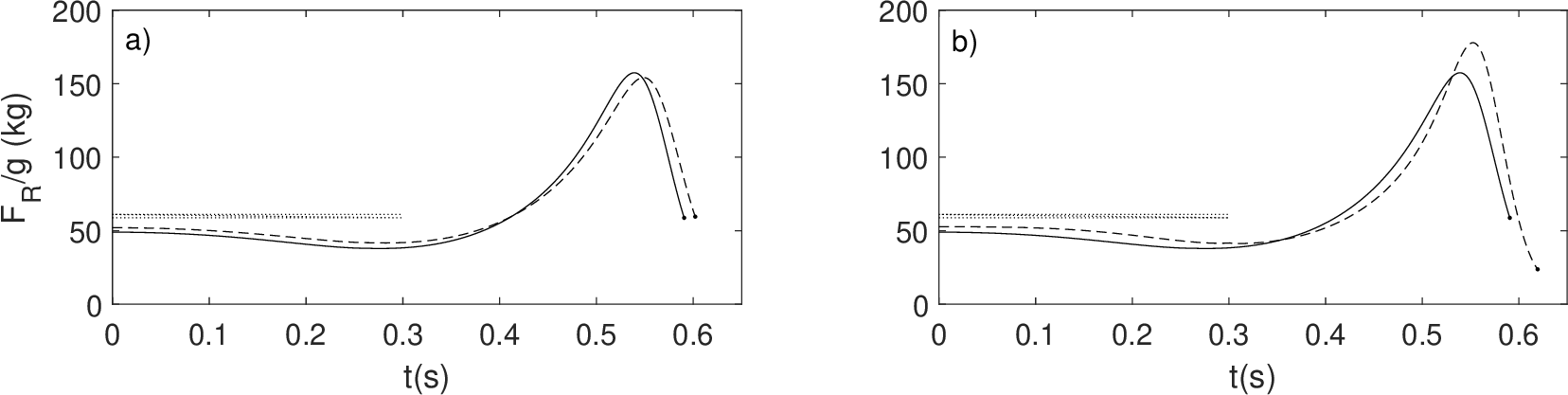}	
		\caption{Forces on shafts for throwing arms measured in units of gravity~$g$. 
		Self consistent result with friction losses.
		a) Full curve, experimental throwing arm. Broken curve, modified arm.
		b) Full curve, experimental throwing arm, same as in a). Broken curve, modified arm and short~$L_3$.
		The total masses of counterweight and throwing arms are shown by horizontal dotted lines.}
		\label{fig:ForcesLosses}
\end{figure}

\Fref{fig:ForcesLosses}a shows time-dependent forces on the shafts for the experimental
and modified throwing arms extending from the start of a shot and until release of the projectile.
The full curve is for the experimental throwing arm, and the broken curve for the replacement.
The curves do not differ much so the dynamics is modified only little as expected because of the
unchanged moment of inertia.
\Fref{fig:ForcesLosses}b shows the reaction force with the shorter~$L_3$ (broken curve),
which is seen to make the shot a little more violent and a little slower, but not much.
\subsection{Scaled trebuchets and losses}
Parameters of the unscaled engine are listed in the first row of~\tref{tab:Param_Alt2}, 
and the next show designs of three larger engines derived by scaling.
\begin{table}[htb]\footnotesize
	\center
		\begin{tabular}{cccccc|ccc|cc}
			\multicolumn{6}{c}{Throwing arm} & \multicolumn{3}{|c|}{} & \multicolumn{2}{|c}{Shafts} 	\\
			$L$ 	& $L_{CM}$	& $m_b$ & $I_B$ 	& $D$		& $S_{\mathrm{arm}}$	& $L_3$	& $M$ 	& $m$ 	
			& $R_R=R_H$ & $L_R=L_H=2D$ \\
			m 		& m 				& kg		& kgm$^2$	& cm 		& \% 				& m 		& kg		& kg		& cm 				& cm 				\\\hline
			1.225	& 0.36   		&	7.21	& 1.85  	&	9.7		& 0.006 		& 0.334	& 53.9	& 0.717	&	1.4   		& 19.4			\\
			4			& 1.18   		&	114		& 312  		&	21.3	& 0.03 			& 1.02 	& 853 	& 11.3	&	4.57 			& 42.6			\\	
			7			& 2.07   		&	421		& 3525  	&	30.9	& 0.06 			& 1.79 	& 3147 	& 41.8	&	8.00 			& 61.9			\\	
			10		& 2.96 			&	968		& 16540  	&	39.2	& 0.10 			& 2.55 	& 7232	& 96.2	&	11.4			& 78.5							
		\end{tabular}
	\caption{Scaled engines:
	Throwing arms and strains~$S_{\mathrm{arm}}$.
	Arms~$L_3$, and masses~$M$ and~$m$ for counterweight and projectile, respectively.
	$S_{\mathrm{shaft}}=0.19\%$. 
	$Q/\D U$=9.50\%.}
	\label{tab:Param_Alt2}
\end{table}
This leaves linear proportions constant at the values~$L_1/L_2=3.90$,~$L_3/L_2=1.34$ 
and~$L_4/L_1=0.89$.
As mentioned earlier, it is also assumed for simplicity that the shafts for the throwing arm 
and the counterweight are identical.
The shafts have lengths~$2D$ and the radii are adjusted such that their estimated strains 
are less that~0.2\%.
The scaling leaves the relative loss of mechanical energy to sliding friction constant at~9.50\%,
and the reaction forces and angular speeds at hinge and fulcrum are such that the losses at each 
shaft are about equal,~4.45\% at fulcrum (see~\tref{tab:Param_Alt1}) and~5.05\% at hinge.

\begin{figure}[htb]
	\centering	
	\includegraphics[width=0.30\textwidth]{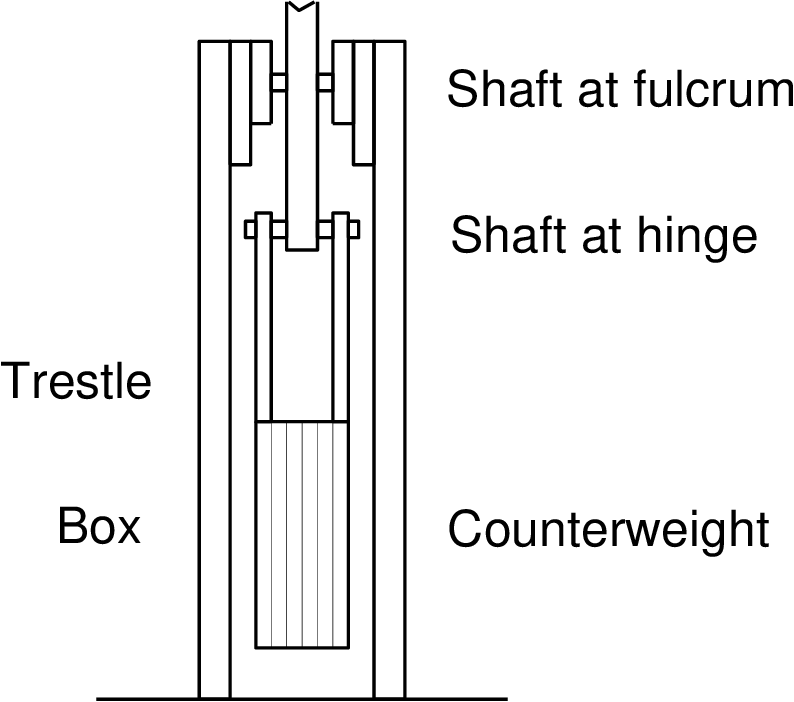}	
		\caption{Trebuchet at rest in final position.
		The shafts for throwing arm and counterweight are seen.}
		\label{fig:Shafts}
\end{figure}

The support of the throwing arm by the trestle and the attachment of the counterweight 
to the hinge at the short end of the throwing arm are illustrated in~\fref{fig:Shafts}.
The trestle is seen to be modified near the fulcrum to support a short shaft, and
the two shafts have equal lengths.  
There is also sufficient room around the counterweight to allow free swing.
\subsection{Ranges, kinetic energies and efficiencies}
Ranges~$R_v$, kinetic energies at a horizontal target~$T_v$ and efficiences~$\en$ 
are listed in the first columns of~\tref{tab:Capacities} for the ideal case without 
loss of mechanical energy.
\begin{table}[htb]\footnotesize
	\center
		\begin{tabular}{c|ccc|ccccc|cc}
			& \multicolumn{3}{|c|}{Ideal} & \multicolumn{5}{|c}{Sliding friction} 	
			& \multicolumn{2}{|c}{Aerodynamic drag} 	\\
			$L$ 	& $R_v$	& $T_v$ & $\en$	& $Q$		& $R_{vf}$	& $T_{vf}$	& $\D T$	& $Q\cdot\en$	& $R_a$	& $T_a$ 		\\
			m 		& m 		& kJ		& \%		& kJ		& m 				& kJ				& kJ			& kJ					& m				& kJ 			\\\hline
			1.225	& 41.3  & 0.158	& 80.1	& 0.019	& 37.7			&	0.146			&	0.012		&	0.015				& 36.0 		& 0.133  	\\
			4			& 135 	& 8.19 	& 80.1	& 0.97 	& 123 			&	7.51			&	0.68		&	0.78				& 116   	& 6.65 		\\	
			7			& 236   & 52.9 	& 80.1	& 6.28	& 216 			&	48.6			&	4.3			&	5.0					& 202   	& 42.4 		\\	
			10		& 337   & 173   & 80.0	& 20.6	& 308				&	159				&	14			&	17					& 286   	& 137						
		\end{tabular}
	\caption{Capacities~$(R,T)$ for the designs in~\tref{tab:Param_Alt2}.
	Ideal vacuum values and efficiencies. 
	Engine losses~$Q$ and vacuum values with losses, losses of energy~$\D T$ 
	and estimates~$\D T\simeq Q\en$.
	Values in air.}
	\label{tab:Capacities}
\end{table}
The larger engines have ranges that could make them interesting as siege weapons, 
and the kinetic energies of the projectiles are also quite large.
Both quantities could be somewhat improved with a better design that optimizes 
the engine and increases the efficiency beyond~90\%,~see~\cite{ref:EH}.
The loss of mechanical energy at the time of release of the projectile is~$Q$, 
and it scales like~$MgL$.
The loss lowers the vacuum range to~$R_{vf}$ and the energy at target to~$T_{vf}$,
and on the assumption that it is distributed on projectile and engine like the 
available mechanical energy, the projectile loses~$\Q\en$.
\Tref{tab:Capacities} shows that~$Q\en$ overestimates the actual loss~$\D T$ by~$\simeq20\%$.
When aerodynamic drag along the ballistic trajectory is taken into account, range
and energy are lowered further as seen in the last two columns of~\tref{tab:Capacities}. 

The heat~$Q$ is generated from a shot is fired and until release of the projectile,
but the engine still possesses mechanical energy at that time, 
and this energy is also dissipated as heat, but now over several oscillation periods 
of throwing arm and counterweight.
For the largest design in~\tref{tab:Capacities} the time until release is~1.7s 
and the heat generated by then is~20.6kJ.
The mechanical energy left in the engine at release is~$\simeq$30kJ so the total
loss to sliding friction is~$\simeq$50kJ from a shot is initiated and until the engine
finally stalls.
Half of this is dissipated after~2.6s,~90\% after~17s and~99\% after~31s. 
The heat is distributed about equally on the four bearings as we have seen, 
and most is generated in less than three seconds. 
The temperature therefore increases sharply, but is not easily calculated.
For reference, the combustion heat for~$\simeq1.5$cm$^3$ of gasoline is close to~50kJ.

More detail on relative reductions of ranges and kinetic energies at target are given 
in~\tref{tab:RelLoss}.
The reduction of range by sliding friction amounts to almost~9\% 
and the further reduction by air drag towards target is a little smaller, 
but increasing with engine size, such that the full reduction relative to the ideal
range without losses range from~13\% to~15\%.
\begin{table}[htb]\footnotesize
	\center
		\begin{tabular}{c|c|c|c|c|c|c|c}
			Throwing arm	& \multicolumn{3}{|c|}{Range reduction} & 
			\multicolumn{3}{|c|}{Energy reduction} & Efficiency	\\\hline
			$L$ 	& Friction 	& Air drag path	& Total	& Friction 	& Air drag path	& Total	& Friction 	\\
			m 		& \% 				& \%				& \%		& \%				& \%				& \%		& \%				\\\hline
			1.225	& 8.7     	& 4.5 			& 13		& 7.6     	& 8.9 			& 16		&	73.0			\\
			4			& 8.9     	& 5.7 			& 14 		& 8.3 			& 11				& 19		&	73.6			\\	
			7			& 8.5    		& 6.5 			& 14 		& 8.1 			& 13 				& 20		&	73.5			\\	
			10		& 8.6     	& 7.1 			& 15 		& 8.1 			& 14				& 21		&	73.3				
		\end{tabular}
	\caption{Reductions of range and kinetic energy at target by friction and air drag.}
	\label{tab:RelLoss}
\end{table}
The calculation of air drag is done on the assumption that the projectiles made of stone 
have indentations from fabrication or naturally, and an assumed drag coefficient~$C$ of~0.25.
The reduction of energy by friction is close to~8\% and the further reduction by air drag 
is almost exactly twice the reduction found for range.
The full reduction is therefore larger than for range and varies from~16\% to~21\%.
The ideal efficiency of~80\% for the trebuchets in~\tref{tab:Capacities} is reduced 
to~$\simeq73\%$ by the sliding friction losses within the engine.
Losses due to air drag along the path do not affect the engine efficiency.
\newpage
\section{Summary and conclusions}
Expressions for generalized friction forces, including sliding friction and aerodynamic drag, 
are determined and added to the equations for the internal movement of a trebuchet 
with swinging counterweight.
The equations can be solved iteratively by the use of perturbation theory when the losses of
mechanical energy are small.

Calculated losses are compared with experimental values obtained with
a smaller trebuchet equipped with motion sensors~\cite{ref:EPJ}.
The comparison is satisfactory and shows,
theoretically as well as experimentally, 
that the losses at the bearings for the pivoting 
beam shaft contribute the most, but smaller losses at the hinge for the swinging counterweight 
and from aerodynamic drag on the sling are also present.
The sliding friction at the hinge is relatively small, because the material here is steel
instead of wood and this allows for a small radius of rotation which implies a small 
sliding speed and a short sliding distance. 
From~\tref{tab:Losses}:
The ideal efficiency of the experimental trebuchet in the absence of mechanical energy loss 
is~$\en=80.4\%$, and at the time when the projectile is released, the accumulated losses amount 
to~$6.6\%$ of the engine's available mechanical energy with most of this loss carried by 
the projectile so the efficiency drops to~75.5\%.
The experimental efficiency is~68.8\%.

These results, which are based on experimental evidence, are scaled to full size 
trebuchets made of wood with realistic throwing arms, 
sliding shafts and lengths of counterweight arms.
The largest engine has a throwing arm measuring~10m, its pivoting point is raised
over the base of the engine by~6.8m and the counterpoise weighs~7232kg. 
This is see in~\tref{tab:Param_Alt2} and from~\tref{tab:Capacities}:
The ideal vacuum capacity for this engine is~$R=337$m and~$T=173$kJ, 
and the ideal efficiency is~$\en\simeq80\%$. 
Sliding friction reduces the vacuum capacity to~$R=308$m and~$T=159$kJ, 
and the efficiency is now~$\en\simeq73\%$.
When air drag along the ballistic path is taken into account, the range on a flat field 
is~$R=286$m and the~96kg projectile arrives at target with a speed of~190km/h 
or the kinetic energy~$T=137$kJ.

\end{document}